\documentclass[conference,10pt,singlespace]{IEEEtran}

\usepackage{graphicx}
\usepackage{amsmath,amssymb}
\usepackage{subfigure}
\usepackage{balance}
\usepackage{algorithm}
\usepackage{algorithmicx}
\usepackage{algpseudocode}
\usepackage[amssymb]{SIunits}

\begin{document}
%\title{Joint Access Point--User Pairing and Linear Precoding Optimization in Distributed MIMO Networks with Limited Data Sharing \\for Fair QoS Experience}

\title{Spatial Resources Optimization in Distributed MIMO Networks with Limited Data Sharing}

%% author names and affiliations
%% use a multiple column layout for up to three different
%% affiliations
%\author{\IEEEauthorblockN{Antonis G. Gotsis and Angeliki Alexiou\\}
%\IEEEauthorblockA{Department of Digital Systems, University of Piraeus, Greece\\
%126 Grigoriou Lampraki Street, Office 501, GR18532, Piraeus, Greece\\
%Email: \{agotsis,alexiou\}@unipi.gr}
%}

\author{\IEEEauthorblockN{Antonis G. Gotsis and Angeliki Alexiou}
\IEEEauthorblockA{Department of Digital Systems, University of Piraeus, Greece\\
%126 Grigoriou Lampraki Street, Office 501, GR18532, Piraeus, Greece\\
Email: \{agotsis,alexiou\}@unipi.gr}
}

%\author{Antonis~G.~Gotsis,~\IEEEmembership{Member,~IEEE,}
%        and~Angeliki~Alexiou,~\IEEEmembership{Member,~IEEE}% <-this % stops a space
%\thanks{A. Gotsis and A. Alexiou are with the Department of Digital Systems, University of Piraeus, Greece, 126 Grigoriou Lampraki Street, Office 501, GR18532, Piraeus, Greece, Email: \{agotsis,alexiou\}@unipi.gr}.% <-this % stops a space
%\thanks{Manuscript received March XX, 2013;}}

%\markboth{Journal of \LaTeX\ Class Files,~Vol.~11, No.~4, December~2012}%
%{Shell \MakeLowercase{\textit{et al.}}: Bare Demo of IEEEtran.cls for Journals}


\maketitle

\begin{abstract}
\boldmath
Wireless access through a large distributed network of low-complexity infrastructure nodes empowered with cooperation and coordination capabilities, is an emerging radio architecture, candidate to deal with the mobile data capacity crunch. In the 3GPP evolutionary path, this is known as the Cloud-RAN paradigm for future radio. In such a complex network, distributed MIMO resources optimization is of paramount importance, in order to achieve capacity scaling. In this paper, we investigate efficient strategies towards optimizing the pairing of access nodes with users as well as linear precoding designs for providing fair QoS experience across the whole network, when data sharing is limited due to complexity and overhead constraints. We propose a method for obtaining the exact optimal spatial resources allocation solution which can be applied in networks of limited scale, as well as an approximation algorithm with bounded polynomial complexity which can be used in larger networks. The particular algorithm outperforms existing user-oriented clustering techniques and achieves quite high quality-of-service levels with reasonable complexity.
\end{abstract}


% \fi
\IEEEpeerreviewmaketitle

\section{Introduction}\label{sec:S1}
With the 4G LTE rollouts gaining momentum all over the world, the so-called ``1000x mobile data challenge"~\cite{NSN11} has initiated intense discussions among the research community for beyond 4G/2020 next generation wireless networks~\cite{3GPP_120045}. A potential solution for dealing with the expected capacity crunch is to massively densify low-cost wireless access network infrastructure and empower it with coordination and cooperation mechanisms, following the recently proposed C-RAN~\cite{CMRI11} or ``Cloud-of-Antennas"~\cite{WeLi12} paradigm. Inspired by the well-known principles of Network MIMO~\cite{KaFo06}, an ultra-dense network could provide linear capacity scaling if the number of infrastructure nodes is at least equal to the number of access terminals and the spatial degrees of freedom are optimally exploited. However, this intelligent spatial resources optimization does not come for free, since the network scalability imposes huge requirements on computational complexity, access nodes synchronization, and overhead for exchanging control and data signaling. First recent efforts have focused on related system requirements and design aspects~\cite{LiWu11}, implementation issues (at least for small-scale networks)~\cite{RaKu12,BaRo13}, transceiver and resources management algorithms~\cite{LiLa12,HoSu13}, as well as overhead reduction techniques~\cite{PeHe12,LiBi12}.

In this work we consider a distributed wireless network consisting of randomly placed infrastructure nodes or simply access points (APs) coordinated by a central entity and serving randomly placed user terminals (UEs).  UEs' data could be fully or partially combined using linear precoding depending on the full (global) channel state information. Similarly to the definition of universal frequency reuse (UFR) in 4G networks, we introduce the concept of ``universal space reuse" (USR) for Distributed MIMO (D-MIMO) networks, in the sense that each AP has access to each UE's traffic, and thus is candidate for transmitting its data. As in UFR where the reuse of frequency resources may be limited due to interference issues~\cite{BoPa09}, in USR space reuse may also be limited due to complexity and overhead issues. In the specific context, we seek for efficient joint AP-UE pairing and linear precoding strategies over the spatial dimension of distributed MIMO setups (that is on a single orthogonal resource unit) when data sharing is limited. From a QoS perspective, we guarantee fair experience in terms of achieved data rate.

\subsection{Related Work}\label{sec:S1a}
The problems of optimum linear precoding for providing fair QoS in single-user and multi-user network MIMO systems have been formulated and exactly solved in \cite{WiEl06} and \cite{ToCo08} through an iterative second order cone programming (SOCP) framework, however these works did not consider data sharing overhead limitations but full cooperation. In \cite{GaKe11} an appealing data sharing cost model was introduced, according to which the overhead was expressed through the number of nodes having access to each UE’s data. However the authors considered the sum-rate QoS maximization problem, which may lead to unfair assigned data rates and also did not provide any solution for the exact optimal precoding desing, but a heuristic algorithm. A similar overhead model was proposed in \cite{ZeGu10} for the sum-power minimization problem in cellular networks with known data rate/SINR requirements, a QoS model more suitable for constant rate voice-like services but not variable-rate data. In addition, the proposed solution approach was based on concepts from sparse optimization (e.g. \cite{CaWa08}) requiring the calculation of a regularization parameter that is generally unknown. Finally, in \cite{ChDr12} the above framework was extended to consider integer activity variables for AP-UE pairings and, proposed a Mixed Integer SOCP (MI-SOCP)  formulation, also involving the unknown regularization parameter. Related simpler heuristic algorithms were also proposed.

\subsection{Contribution - Paper Organization}\label{sec:S1b}
In this work, inspired by the framework of \cite{ChDr12}, we first formulate the fairness-based QoS problem utilizing an MI-SOCP model as well, however, without including any regularization parameter, which can be exactly solved by modern mixed integer conic solvers. Secondly we propose and explore different variations for the data sharing limitation, that is either the overall number or the per-UE number of APs-to-UEs pairings is constrained. Third due to complexity and lack of scalability of the exact integer optimization model, we propose a faster approximation algorithm which iteratively solves a series of SOCP pproblems and is demonstrated to perform quite-high under bounded and predictable computational complexity. The simpler algorithm outperforms baseline schemes which are based on clustering principles of multi-cell MIMO networks (e.g. \cite{NgHu10}). While the exact MI-SOCP solution may be utilized for smaller networks (up to 10x10 in typical computational platforms) due to its high computational requirements, the heuristic may be used for realistic larger scale ultra-dense networks.

The rest of the paper is structured as follows. In Section~\ref{sec:S2} system modeling issues along with the work assumptions are presented. In Section~\ref{sec:S3} the optimal formulation and solution through the mixed integer second order cone programming framework is given, while in Section~\ref{sec:S4} the proposed faster approximation algorithm is explained. In Section~\ref{sec:S5} simulation results are given while in Section~\ref{sec:S6} the work is concluded and future work directions are stated.

\section{System Modeling and Assumptions}\label{sec:S2}

Let a wireless network comprising $M$ randomly placed distributed access nodes/points (APs) serving $K$ users (UEs) in a coordinated way following the C-RAN/Network MIMO paradigm. A single resource unit (RU) is considered (for example an LTE-A time-frequency resource block), which all APs/UEs can access through spatial multiplexing. Extension to multiple orthogonal resource units is possible by independently applying the system model to each RU separately. The APs have knowledge of the network channel state information (CSI) through over-the-air feedback links with the UEs, and full/partial access to UEs' data through backhaul links to facilitate linear precoding. From a quality-of-service perspective, we aim at providing fair UE experience following the LTE-Advanced service requirements. This is achieved by guaranteeing common data rate per UE or equivalently through a Shannon-based PHY-abstraction approach \cite[Sec.10.4]{HoTo11} common signal to interference plus noise ratio (SINR). In this sense, we seek for the maximum common SINR, let $t^*$, and how to realize it, that is the optimal complex-valued precoding matrix ${\bf{W}} \in {{\Bbb C}^{M \times K}}$ to apply to UEs' data.

For the partial data sharing case (imposed by backhaul or complexity limitations) each $\left\{ {m,k} \right\}$ element of the precoding matrix is actually a semi-continuous complex quantity since it could also be zero when the corresponding AP$_m$ does not have access to UE$_k$ data, that is the specific AP-UE pair is not active. Therefore, as in \cite{ChDr12}, we also define an auxiliary binary-valued AP-UE pairing matrix ${\bf{\alpha }} \in {\left\{ {0,1} \right\}^{M \times K}}$, for which the arbitrary $\alpha_{mk}$ element is $1$ if the $m^{\text{th}}$ AP and the $k^{\text{th}}$ UE are paired else is $0$. The latter matrix $\bf{\alpha}$ can be seen as a virtual connectivity matrix, and an all-ones matrix corresponds to the full data sharing scenario. At the other extreme, each column of $\bf{\alpha}$ has exactly one non-zero element, in order for each UE to be served by at least one AP and assigned a non-zero data rate, due to the fairness-based QoS profile.

\begin{figure}
\centering
\includegraphics[scale=0.38]{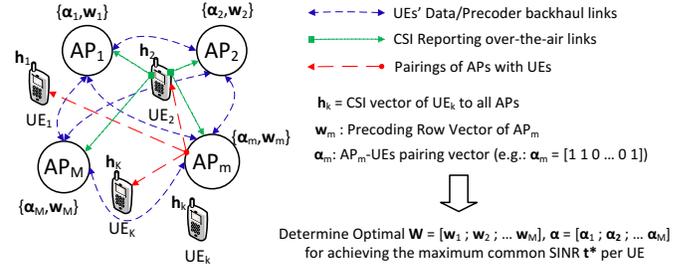}
\caption{System Model}
\label{fig:DMIMO}
\end{figure}

Perfect Network CSI (measured and reported by the UEs) is assumed which corresponds to the normalized complex channel matrix ${\bf{H}} \in {{\Bbb C}^{K \times M}}$ and includes path-loss, shadowing, small-scale fading as well as the effect of noise. The latter is applied by normalizing the matrix with the noise amplitude $\sqrt {{N_0}\Delta f} $, where $N_0$ is the noise power density and $\Delta f$ the resource unit transmission bandwidth. The matrix $\bf{H}$ is structured by the per-UE row vectors ${{\bf{h}}_k} \in {{\Bbb C}^{M \times 1}}$, and accordingly the precoder matrix $\bf{W}$ by stacking up the per-UE column vectors $\bf{w}_k$ or equivalently the per-AP row vectors $\bf{w}_m$. The offered per-UE QoS is expressed through the achieved SINR $\gamma_k$ which is simply given by ${\gamma _k} = \dfrac{{{{\left\| {{{\bf{h}}_k}{{\bf{w}}_k}} \right\|}^2}}}{{1 + \sum\limits_{i \ne k} {{{\left\| {{{\bf{h}}_k}{{\bf{w}}_i}} \right\|}^2}} }}$. System operation is also characterized by a maximum power budget $P_\text{max}^{\text{AP}}$ on the transmission (Tx) power per AP $P_m$, where ${P_m} = \sum\limits_k {{{\left\| {{w_{mk}}} \right\|}^2}} $. In addition, an upper bound on the overall or on the per-UE basis number of pairings or ``virtual connections" among APs and UEs is imposed. For the former case the number of pairings $\sum\limits_k {\sum\limits_m {{a_{mk}}} } $ is simply upper bounded by an integer number $b_{tot}$ while for the latter each UE's pairings number $\sum\limits_m {{a_{mk}}} $ is upper bounded by ${\raise0.7ex\hbox{${b_{{tot}}}$} \!\mathord{\left/
 {\vphantom {{b_{{tot}}} K}}\right.\kern-\nulldelimiterspace}
\!\lower0.7ex\hbox{$K$}}$.
An illustration of the considered system model is given in Figure~\ref{fig:DMIMO} where an AP-UE pairing example is also given.

\section{Optimal Solution}\label{sec:S3}
We first cast the particular QoS provision scenario as an optimization problem and provide a method to obtain the exact maximum common rate per UE along with the optimal AP-UE pairing and Linear Precoding solutions. Towards this purpose, we follow the work of \cite{ChDr12}, however our approach differs from the previous one in a threefold way; first we consider the fairness problem, while in \cite{ChDr12} the minimum sum-power minimization for given SINR constraints was examined; secondly, the proposed formulation included a regularization parameter which is unknown, while our model is related only to the actual system parameters; thirdly we introduce new data sharing constraints (per-UE), which have practical interest for the system designer. The mathematical programming formulation is given in Eq.\eqref{eq:PForig}.

\begin{subequations}\label{eq:PForig}
\begin{gather}
\mathop {\max }\limits_{t,{\bf{W}},{\bf{\alpha }}} t{\rm{ }} \hfill \label{eq:PForig_a}\\
{\text {subject to}} \hfill \nonumber \\
{\gamma _k} \ge t,\forall k{\text{, where }}{\gamma _k} = \frac{{{{\left\| {{{\bf{h}}_k}{{\bf{w}}_k}} \right\|}^2}}}{{1 + \sum\limits_{i \ne k} {{{\left\| {{{\bf{h}}_k}{{\bf{w}}_i}} \right\|}^2}} }} \hfill \label{eq:PForig_b} \\
{P_m} \le P_{\max }^{AP},\forall m{\text{, where }}{P_m} = \sum\limits_k {{{\left\| {{w_{mk}}} \right\|}^2}} \hfill \label{eq:PForig_c} \\
\left\| {{w_{mk}}} \right\| \le {a_{mk}} \cdot \sqrt {P_{\max }^{AP}} ,\forall m{\rm{,}}\forall k \hfill \label{eq:PForig_d} \\
\sum\limits_k {\sum\limits_m {{a_{mk}}} }  \le {b_{tot}}{\text{ or }}\sum\limits_m {{a_{mk}}}  \le {\raise0.7ex\hbox{${{b_{tot}}}$} \!\mathord{\left/
 {\vphantom {{{b_{tot}}} K}}\right.\kern-\nulldelimiterspace}
\!\lower0.7ex\hbox{$K$}},\forall k \hfill \label{eq:PForig_e}
\end{gather}
\end{subequations}

By simple inspection this problem is a Mixed Integer Nonlinear Programming model which is extremely hard to solve \cite{LeLe12}. However, for a specific given minimum SINR constraint $t_i$ and after some manipulations as in \cite{KaFo06,ToCo08}, each \eqref{eq:PForig_b} constraint can be rewritten in two expressions $\left\| {\left[ {\begin{array}{*{20}{c}}
1\\
{{{\left( {{\bf{h}}{}_k{\bf{W}}} \right)}^H}}
\end{array}} \right]} \right\|
\le \sqrt {1 + \dfrac{1}{{{t_i}}}}  \cdot \Re \left\{ {{{\bf{h}}_k}{{\bf{w}}_k}} \right\}$, $\Im \left\{ {{{\bf{h}}_k}{{\bf{w}}_k}} \right\} = 0$, which are conic (a norm left-hand-side lower than or equal to an affine right-hand-side~\cite{LoVa98}) and linear respectively. Constraints \eqref{eq:PForig_c},\eqref{eq:PForig_d} are also conic and \eqref{eq:PForig_e} linear ones, therefore for a given SINR target, the problem is a Mixed-Integer Second-Order Conic Program (MI-SOCP), which can be exactly solved by modern solvers like MOSEK~\cite{MOSEK} and CPLEX~\cite{CPLEX}. The respective formulation is given in Eq.\eqref{eq:PFmisocp}.

\begin{equation}\label{eq:PFmisocp}
\begin{array}{l}
{\text{find }}{\bf{W}},{\bf{\alpha }}{ \text{ s.t. } }\left\| {\left[ {\begin{array}{*{20}{c}}
1\\
{{{\left( {{\bf{h}}{}_k{\bf{W}}} \right)}^H}}
\end{array}} \right]} \right\| \le \sqrt {1 + \dfrac{1}{{{t_i}}}} \Re \left\{ {{{\bf{h}}_k}{{\bf{w}}_k}} \right\},\forall k,\\
\Im \left\{ {{{\bf{h}}_k}{{\bf{w}}_k}} \right\} = 0,\forall k \text{ and \eqref{eq:PForig_c}, \eqref{eq:PForig_d}, \eqref{eq:PForig_e}}
\end{array}
\end{equation}

With respect to locating the maximum common SINR, we could apply a bisection search~\cite{BoVa04,ToCo08}, where at each step the particular feasibility MI-SOCP problem is solved (equivalently instead of feasibility problems, we may set the \eqref{eq:PForig_e} expression as the objective function as well). Therefore, the exact optimal solution could be obtained by this iterative {MI-SOCP} procedure, which is given in pseudocode format in Alg.\ref{alg:Optimal} for convenience. Finally note that the left \eqref{eq:PForig_e} constraint expressions represent the most flexible pairing allocation, while the rightmost ones the per-UE data sharing limitations, so the achieved QoS levels for the former case is the upper bound for the latter or any other data sharing constraint.

\begin{algorithm}
\small
\caption {Bisection Search for Maximum Common SINR}
\begin{algorithmic}[1]
\State Given $t_l \le t \le {t_{u}}$, convergence tolerance $\epsilon$
\Repeat
    \State $t_i \leftarrow {{\left( {{t_l} + {t_u}} \right)} \mathord{\left/
 {\vphantom {{\left( {{t_l} + {t_u}} \right)} 2}} \right.
 \kern-\nulldelimiterspace} 2};$
    \State Solve the MI-SOCP feasibility problem \eqref{eq:PFmisocp};
    \State If a feasible solution is found $t_u \leftarrow t_i$ else $t_l \leftarrow t_i$;
\Until {$\left| {{t_u} - {t_l}} \right| \le \varepsilon$}.
\end{algorithmic}
\label{alg:Optimal}
\end{algorithm}

\section{A Faster Approximation Algorithm}\label{sec:S4}

\subsection{Motivation}\label{sec:S4a}
Although the problem has been formulated as an attractive mathematical program, the complexity remains extremely high, due to the involved integer optimization variables needed for defining the 0-1 pairing matrix. It is known that the worst-case solution complexity of such problems scales exponentially with the problem size and the time to solve it can not be bounded (see for example~\cite{WoNe99}). Therefore the exact solution approach is realistic only for small-scale networks. In order to deal with larger problem dimensions we can not rely on this model, but we need a simpler lower-complexity algorithm, still achieving high quality performance levels.

\subsection{Algorithm Description and Complexity}\label{sec:4b}
The complexity burden of the original formulation lies in the binary pairing matrix. For a given pairing matrix $\bf{\alpha}$, the problem solved at each iteration of the bisection search becomes a simple SOCP (see Eq.\eqref{eq:PFprecod}) which is convex, and can be solved exactly and efficiently (in terms of execution time) for large dimensions through available convex optimization solvers like SDPT3~\cite{SDPT3}.

\begin{equation}\label{eq:PFprecod}
\begin{array}{l}
{\text{find }}{\bf{W}}{ \text{ s.t. } }\left\| {\left[ {\begin{array}{*{20}{c}}
1\\
{{{\left( {{\bf{h}}{}_k{\bf{W}}} \right)}^H}}
\end{array}} \right]} \right\| \le \sqrt {1 + \dfrac{1}{{{t_i}}}}  \cdot \Re \left\{ {{{\bf{h}}_k}{{\bf{w}}_k}} \right\},\forall k,\\
\Im \left\{ {{{\bf{h}}_k}{{\bf{w}}_k}} \right\} = 0,\forall k,\sum\limits_k {{{\left\| {{w_{mk}}} \right\|}^2}}  \le P_{\max }^{AP},\forall m{\rm{,}}\\
{w_{mk}} = 0,\forall m,k{\text{ having }}{a_{mk}} = 0
\end{array}
\end{equation}
Based on this observation, we propose a ``greedy-like" procedure to approach the data sharing constrained pairing matrix, for which at each step a series (due to the bisection search) of conic feasibility or optimization problems is solved, hence the pairing matrix is approached from the exterior of the feasible space~\cite{Ra96}. In particular, we begin with an all-ones pairing matrix (which corresponds to the full data sharing case or in other words to the relaxation of the \eqref{eq:PForig_e} constraints) and gradually (one-by-one) zero specific elements of it, until a matrix fulfilling the \eqref{eq:PForig_e} constraint(s) is acquired. The suboptimality of this approach stems from the selection of the element to zero at each step of the greedy algorithm (since a heuristic criterion is applied) as well as the gradual (and not the ``single-shot") selection procedure. The advantage of this approach is that it has a bounded and predictable computational complexity and is robust, since it is based on simple heuristic zeroing criterion.

As far as the selection of which AP-UE pairing to exclude from our solution at each step, we first explain what is the effect of zeroing an arbitrary $w_{mk}$ element. Based on the expression for the SINR per UE given in Eq.\eqref{eq:PForig_b} we observe that both the received power for the particular UE and the interference power to all other UEs are affected. Therefore, in order to consider both quantities and balance both effects, we define our selection criterion as the ratio of the new received power to the interference power and select the combination which maximizes this ratio. The complete algorithmic steps are given below.
\begin{algorithm}
\caption {Greedy-like Approximation Algorithm}
\small
\begin{description}
  \item[\textbf{Step 1}] \hspace{0pt} \textit{(Initial Full Data Sharing Solution)}\\
  Run bisection search on Problem~\eqref{eq:PFprecod} for full data sharing conditions and acquire an initial optimal precoder.
  \item[\textbf{Step 2}] \hspace{0pt} \textit{(Greedy Selection of Pairing Matrix Element to Zero)}\\
  Based on the current precoding matrix calculate for each possible AP-UE $\left\{ {m,k} \right\}$ combination the following two quantities if this element is zeroed, that is, if $w_{mk}=0$:\\
  i) Received Power for the $k^{\text{th}}$ UE: $R{P_{mk}} = {\left\| {{h_k}{w_k}} \right\|^2}$\\
  ii) Interference Power to rest UEs: ${I_{mk}} = \sum\limits_{i \ne k} {\sum\limits_{j \ne i} {{{\left\| {{h_i}{w_j}} \right\|}^2}} } $\\
  Then, select the combination $\left\{ {m^*,k^*} \right\}$ that maximizes the ratio of $R{P_{mk}}/{I_{mk}}$, and re-run bisection search for problem \eqref{eq:PFprecod} with the selected pairing matrix element forced to be zero (in addition to previously zeroed elements) and acquire a new precoder matrix.
  \item[\textbf{Step 3}] \hspace{0pt} \textit{(Data Sharing Constraint Check)}\\
  Check if the number of non-zero elements of the pairing (or equivalently precoder) matrix exceeds the \eqref{eq:PForig_e} data sharing constraint. If yes go to Step 2, else terminate algorithm.
\end{description}
\label{alg:IterMISOCP}
\end{algorithm}

As for the estimation of worst-case computational complexity order for the proposed greedy-like algorithm, we first notice that there are two iteration levels, an outer for greedily reducing the number of AP-UE pairings and an inner for reoptimizing the precoder matrix. Regarding the outer procedure, it reduces the AP-UE pairings from $MK$ at the initialization phase to the required number set by the data sharing constraint, $b_{tot}$, hence for the stricter limitation case requires $MK - b_{tot} = MK - K  \approx MK$ iterations. For the inner procedure, first the best AP-UE pairing is selected among the $MK$ possible combinations, with complexity $MK$. Then, a bisection search for the common SINR optimization is employed, which needs exactly $\left\lceil {{{\log }_2}\left[ {{{\left( {{t_u} - {t_l}} \right)} \mathord{\left/
 {\vphantom {{\left( {{t_u} - {t_l}} \right)} \varepsilon }} \right.
 \kern-\nulldelimiterspace} \varepsilon }} \right]} \right\rceil $ \cite[Sec.4.2.5]{BoVa04} iterations (for typical parameters selection given in Section~\ref{sec:S5}, there occur 20 iterations per bisection search). Finally, at each iteration of the bisection search a SOCP is solved. According to \cite{LoVa98} for a second order conic program with $n$ variables and $i$ conic constraints with dimension $n_i$ each, complexity scales by ${\rm O}\left( {n\sum\limits_i {{n_i}} } \right)$. Our formulation given in \eqref{eq:PFprecod} involves $MK$ variables (precoding matrix elements), $K$ conic constraints of dimension $K$ each for guaranteeing the minimum SINR per UE, and $M$ conic constraints of dimension $K$ each for bounding the Tx power per AP. Hence, the SOCP sub-step scales by ${\rm O}\left( {MK \cdot \left( {{K^2} + MK} \right)} \right) = {\rm O}\left( {MK \cdot \left( {{K^2} + MK} \right)} \right) \approx {\rm O}\left( {{{\left( {MK} \right)}^2}} \right)$. Combing the above calculations and after some manipulations, we get a 3rd-order polynomial ${\rm O}\left( {{{\left( {MK} \right)}^3}} \right)$ complexity order with respect to the network size, contrary to exponential $2^{MK}$ of the exact optimal\footnote{Note that exhaustive search approach is also unaffordable due to the excessive number of AP-UE pairing combinations. For example for a 5x5 network there exist 155 potential AP-UE combinations, and for each one the optimal linear precoding should be calculated as well.}.

\section{Results and Discussion}\label{sec:S5}

\subsection{Simulation Setup}
A series of Monte-Carlo simulation experiments is performed in order to acquire insights on the optimal AP-UE pairing patterns for the two data sharing limitation strategies (overall/per-UE) and the achieved performance levels of the faster greedy-like algorithm. Each computational experiment simulates a distributed wireless access network, comprising $M$=6 APs and $K$=6 UEs (non-symmetrical MIMO networks are also fully supported by our framework). Each transmission occupies a single resource unit of $200$~\kilo\hertz. Network MIMO channel matrices are generated based on the WinnerII model~\cite{WINII07a} and simulate both large-scale (path-loss, shadowing) and small-scale (fading) effects, while thermal noise with density $N_0=-174$ \deci\bel\milli/\hertz~corrupts received signals. Each AP power is upper bounded by $P_{\text{max}}^{\text{AP}}=1$ \watt, and APs/UEs are randomly dropped over a square area with side length ($R_\text{max}$) $1000 \text{ or } 250$ \metre. The reference SNR per link (averaged over all AP-UE combinations and computed assuming uniform power allocation per UE and maximum power per AP) is 16~\deci\bel~for the sparser and 26~\deci\bel~for the denser setup.

Regarding the bisection search, we bound the target SINR as ${t_l} = 0,{t_u} = {10^4}$ and set the termination tolerance to $\varepsilon  = 0.01$. As far as the optimization problems, we use {CVX}~\cite{CVX} to model the optimization formulations in {MATLAB} along with the {SDPT3}~\cite{SDPT3} and {MOSEK}~\cite{MOSEK} packages for obtaining the solution of conic and mixed integer conic instances. Regarding the mixed-integer problems we set an execution upper bound of $900$~\second. Network Rate is calculated by a PHY abstraction (see \cite[Sec.10.4]{HoTo11}), is evenly divided to the UEs due to fair QoS profiling, and is averaged over 20 independent setups.

\subsection{Simulation Results \& Comments}
\begin{figure}

\subfigure[Data Sharing Limitation Scenario \#1]{
\includegraphics[width=0.45\textwidth]{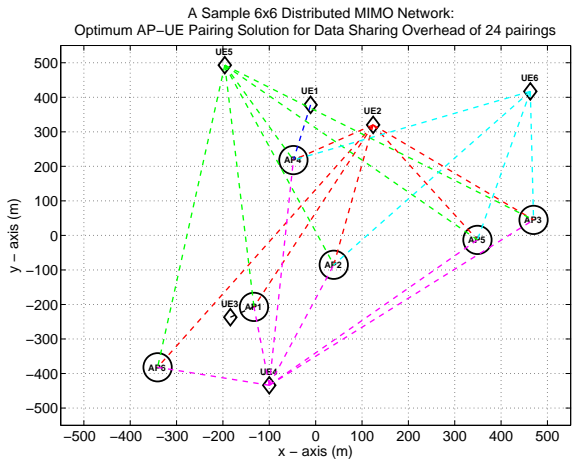}
\label{fig:SystemSnapshots_1} }
\subfigure[Data Sharing Limitation Scenario \#2]{
\includegraphics[width=0.45\textwidth]{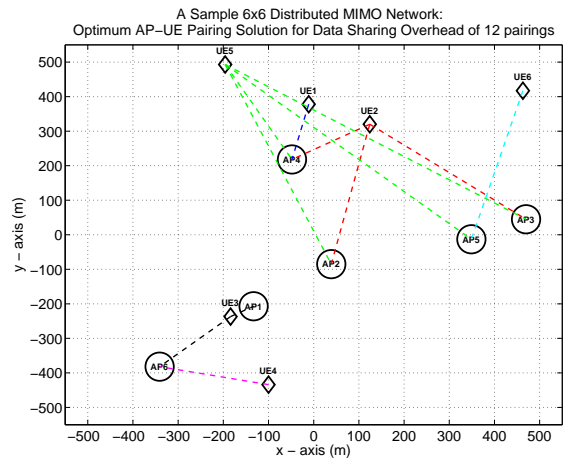}
\label{fig:SystemSnapshots_2} }

\caption[]{Optimum AP-UE pairings for a random network snapshot and two data sharing limitation scenarios}
\label{fig:SystemSnapshots}
\end{figure}

We first summarize the implemented schemes:
\begin{itemize}
  \item Optimal Joint AP-UE Pairing and Linear Precoding with overall (``\textbf{OPT}") and per-UE (``\textbf{OPT-perUE}") data sharing constraints (Alg.\ref{alg:Optimal}-Section~\ref{sec:S3}).
  \item The proposed faster greedy-like approximation algorithm (``\textbf{APPROX}"), refer to Alg.\ref{alg:IterMISOCP}-Section~\ref{sec:S4}.
  \item Two disjoint 2-step AP-UE pairing \& linear precoding baseline ``user-clustering" algorithms (``\textbf{CLUST-1/2}"). At the 1st step each UE chooses its
      best ${{{b_{tot}}} \mathord{\left/
 {\vphantom {{{b_{tot}}} K}} \right.
 \kern-\nulldelimiterspace} K}$ APs to be served based on either a ``maximum channel gain" or ``maximum interfering UEs" criterion (see \cite{NgHu10} for a description of the corresponding approaches) and then independent optimal linear precoding is performed.
 \item A scheme based on random AP-UE pairing association combined with a subsequent optimal linear precoding (``\textbf{RANDOM}").
\end{itemize}
In Figure~\ref{fig:SystemSnapshots} the optimal active AP-UE pairings acquired through Algorithm~\ref{alg:Optimal} are graphically illustrated for two data sharing constraint scenarios (a ``lighter" and a ``tighter"). It is observed that the operation mode for the whole network is a mix of multi-user MIMO,MISO and SISO transmission strategies, and that for tighter limitations (see Figure~\ref{fig:SystemSnapshots_2}) the network is decoupled (partitioned) into smaller subnetworks. Also, on a per-UE basis, an unequal number of serving APs is assigned, demonstrating that the balanced clustering approach is suboptimal.

\begin{figure}
\centering
\subfigure[Sparser Network]{
\includegraphics[scale=0.39]{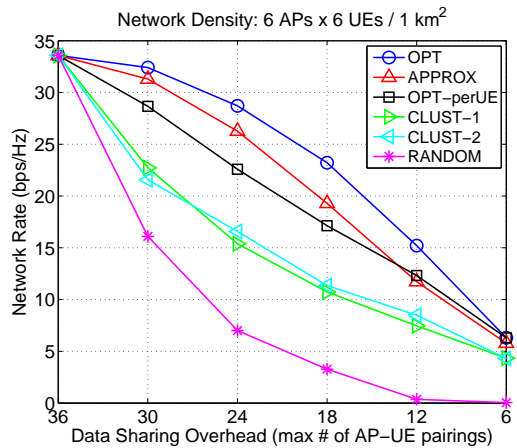}
\label{fig:RESULTS_1000}
}
\subfigure[Denser Network]{
\includegraphics[scale=0.39]{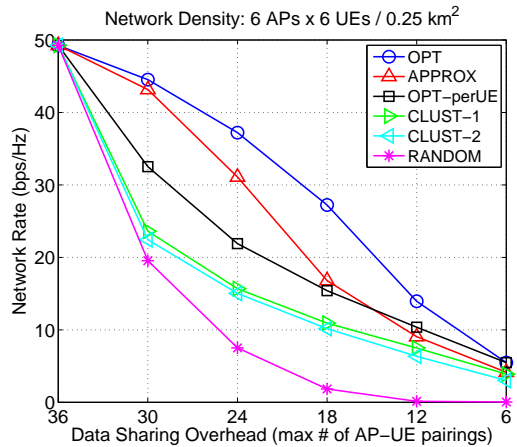}
\label{fig:RESULTS_500}
}
\caption[]{Comparative performance results for data sharing constrained common rate optimization}
\label{fig:RESULTS}
\end{figure}

Figure~\ref{fig:RESULTS} depicts the comparative achieved (common) rate performance for the examined schemes regarding two different network density setups. We first remark (by comparing ``OPT-perUE" with ``OPT") that when we impose data sharing constraints on a per-UE basis, namely we require that all the UEs should have the same number of serving APs, a performance penalty is induced. In other words, the ideal AP-UE association suggests an unbalanced pairing for the UEs (refer to the previous empirical observations as well). Actually, UEs which were very close to APs were allocated smaller number of APs with low Tx power, contrary to other UEs that did not have a “high quality” AP near to them, and were allocated a larger number of APs. The suboptimality levels of the ``perUE" limitation scenario goes up to 27\% for the sparser setup and 46\% for the denser one.

Secondly, the two baseline clustering algorithms (``CLUST-1/2") experience a large optimality gap from their upper bound (``OPT-perUE") since they are based on disjoint AP-UE pairing and linear precoding. Therefore, a joint pairing-precoding optimization is necessary for achieving high performance levels. Thirdly, the proposed greedy-like scheme exhibits acceptable performance with reasonable complexity. It even outperforms (or is very close to) the ``perUE" upper bound, since it does not perform balanced pairing among UEs, but a very flexible approach is applied at each step of locating the pairs to remove. The proposed algorithm achieves on average the 90\% (Fig.\ref{fig:RESULTS_1000}) and 80\% (Fig.\ref{fig:RESULTS_500}) and for the worst-case scenario the 78\% (Fig.\ref{fig:RESULTS_1000}) and 65\% (Fig.\ref{fig:RESULTS_500}) of the optimal network rate for the different network density setups. We believe that for denser network setups and more challenging data sharing limitations, the proposed greedy algorithm performance deteriorates due to the large number of necessary heuristic searches ($MK-b_{tot})$ and the increase in experienced interference levels, which are difficult to be handled by our simple pairing removal criterion. The development of other greedy schemes with more sophisticated criteria is left for future work, however the particular scheme already achieves to a sufficient extent the merits of joint pairing and precoding optimization, contrary to the disjoint baseline clustering schemes.

\section{Conclusion \& Future Work}\label{sec:S6}
In this paper, the problem of joint design of access points and users pairing as well as linear precoding for distributed MIMO networks was thoroughly examined. Solutions for networks with finite data sharing capabilities and fairly assigned QoS levels across the users, were proposed. First, a method for obtaining the optimal D-MIMO design has been devised, and second, a faster approximation algorithm with significantly lower complexity but still quite high QoS performance also shown to outperform existing user-clustering schemes, was proposed. The importance of jointly considering the pairing and the precoding sub-procedures to D-MIMO design was demonstrated. Future work includes the improvement of the proposed approximation algorithm, the consideration of other mixed QoS models, and the application of the above concepts in very-large ultra-dense access networks.

\section*{Acknowledgment}
This work has been performed in the context of the ART-COMP PE7(396)\textit{ ``Advanced Radio Access Techniques for Next Generation Cellular NetwOrks: The Multi-Site Coordination Paradigm"} and THALES-INTENTION research projects, conducted within the framework of Operational Program "Education and Lifelong earning", co-financed by the European Social Fund (ESF) and the Greek State.


%\balance
\bibliographystyle{IEEEtran}
\bibliography{ARTCOMP-bib}

%\bibliographystyle{IEEEtran}

\end{document}